\documentclass[aps,twocolumn,groupedaddress,superscriptaddress]{revtex4-1}

\usepackage{graphicx,psfrag,color,hyperref}
\usepackage{amssymb,latexsym,mathrsfs,amsmath,amsthm}
\usepackage[utf8]{inputenc}
\usepackage{subfigure}
\usepackage{float}

\begin{document}
  
\def\be{\begin{equation}}
\def\ee{\end{equation}}
\def\bea{\begin{eqnarray}}
\def\eea{\end{eqnarray}}

\title{Stochastic quantum Zeno-based detection of noise correlations}

\author{Matthias M. Müller}
\affiliation{\mbox{Department of Physics and Astronomy, LENS, and \mbox{QSTAR}, University of Florence,} via G. Sansone 1, I-50019 Sesto Fiorentino, Italy.}

\author{Stefano Gherardini}
\affiliation{\mbox{Department of Physics and Astronomy, LENS, and \mbox{QSTAR}, University of Florence,} via G. Sansone 1, I-50019 Sesto Fiorentino, Italy.}
\affiliation{Department of Information Engineering, \mbox{INFN}, and \mbox{CSDC}, University of Florence, via G. Sansone 1, I-50019 Sesto Fiorentino, Italy.}

\author{Filippo Caruso}
\affiliation{\mbox{Department of Physics and Astronomy, LENS, and \mbox{QSTAR}, University of Florence,} via G. Sansone 1, I-50019 Sesto Fiorentino, Italy.}

\begin{abstract}
A system under constant observation is practically freezed to the measurement subspace. If the system driving is a random classical field, the survival probability of the system in the subspace becomes a random variable described by the Stochastic Quantum Zeno Dynamics (SQZD) formalism. Here, we study the time and ensemble average of this random survival probability and demonstrate how time correlations in the noisy environment determine whether the two averages do coincide or not. These environment time correlations can potentially generate non-Markovian dynamics of the quantum system depending on the structure and energy scale of the system Hamiltonian. We thus propose a way to probe this interesting property of the environment by means of the system survival probability. This will further contribute to the development of new schemes for quantum sensing technologies, where nanodevices may be exploited to image external structures or biological molecules via the surface field they generate.
\end{abstract}

\maketitle

\section*{Introduction}
The dynamical evolution of a quantum system is always influenced by its environment~\cite{Petruccione1,CarusoRMP}. Since one is very often only interested on the system dynamics, the environmental degrees of freedom are traced out and, in the Markovian regime (under the assumption of only very short-lived correlations), this leads to the well-known Kossakowski-Lindblad master equation~\cite{Kossakowski}. As a consequence, this approximation does not take into account all the environment-induced memory effects, which may produce a back flow of information onto the quantum system~\cite{Breuer1,Bylicka1}. However, the environment is usually unknown and very hard to be characterized. Therefore, there is a growing interest in the characterization of the environment according to whether it can generate Markovian or non-Markovian dynamics of the system to which it is coupled, where also the latter category can be subdivided into a full hierarchy of non-Markovianity~\cite{Rivas}. Classical environments exhibiting non-Gaussian fluctuations (i.e. characterized by non-Gaussian probability density functions) can lead to non-Markovian quantum dynamics~\cite{Benedetti1, Benedetti2}. Indeed, non-Gaussian stationary stochastic processes cannot be described only by the mean and the variance of the first order density function, and they represent the natural mathematical tool to characterize the structure of an arbitrary environment.
In this context, the latter can be probed by coupling a (typically small, e.g. one qubit) quantum system of known dynamics to it, and studying the effect of the environment on the system dynamics. Indeed, the very recent idea of the so-called quantum probes is that their fragile properties, as coherence and entanglement, are strongly affected by the environment features and can be used to detect them. Examples of such physical systems are quantum dots, atom chips and nitrogen vacancy centers in diamond where a good control over the system has been proposed and recently achieved~\cite{Taylor, Balasubramanian1, Maze, McGuinness, Hollenberg, Hofferberth, Gierling, Ockeloen, Rossi1}. They can be used to probe environments like biological molecules or surfaces of solid bodies or amorphous materials. On the other side, a number of non-Markovianity measures and witnesses has been proposed, such as geometric measures (i.e. measures based on the geometry of the space of quantum maps), quantities based on the Helstrom matrix (i.e. based on the distinguishability of two states under evolution and observation), or witnesses based on the (non-)monotonicity of entanglement measures~\cite{Rivas}. Most of them, however, rely on a full state
tomography and are thus experimentally difficult to be implemented. An experimentally feasible tool for certain systems is based on the state
distinguishability and the Loschmidt echo~\cite{Haikka}.

Recently, the scenario of stochastic measurement sequences has been proposed~\cite{Shushin1}, and then studied with a particular focus on the probability for the system (survival probability) to remain confined within a given subspace~\cite{Gherardini1,Gherardini2,Mueller2016}. Indeed, when the time interval between two measurements is random, this survival probability becomes a random variable by itself, and it has been shown by large deviation theory~\cite{Ellis1,Touchette1,Dembo1} that it converges to its most probable value, by increasing the number of the measurements performed on the system~\cite{Gherardini1}. When the measurementes become very frequent, the survival probality increases and a stochastic quantum Zeno regime is accessed~\cite{Gherardini1,Mueller2}. It is the stochastic generalization of quantum Zeno dynamics (QZD), where in the limit of infinitely frequent observation the dynamics of a quantum system is freezed to a unidimensional~\cite{Misra1} or multidimensional~\cite{Pascazio1,Smerzi1} subspace of the measurement operator.
QZD has been experimentally realized first with a rubidium Bose–Einstein condensate in a five-level Hilbert space \cite{Schafer1}, and later in a multi-level Rydberg state structure \cite{Signoles1}. Furthermore, a recent theoretical study and experimental demonstration with atom-chips has shown also how different statistical samplings of a randomly-distributed sequence of projective measurements coincides in the quantum Zeno regime, proving an ergodicity hypothesis for randomly perturbed quantum systems~\cite{Gherardini2}. In this regard, the sensitivity of the survival probability to the stochasticity in the time interval between measurements has been properly analyzed by means of the Fisher information~\cite{Mueller2016}.

In this work, we propose a method based on the Stochastic Quantum Zeno Dynamics (SQZD)~\cite{Gherardini1,Mueller2} to detect time correlations in random classical fields. Indeed, we use the SQZD formalism to study a quantum system, subjected to a sequence of equally spaced projective measurements, interacting with an environment modelled by a randomly fluctuating field. Then, the random value of the field leads to a random value of the survival probability in the measurement subspace.
As outline, we first introduce our model of a quantum system coupled to the environment. Then, we review and adapt the SQZD formulation, and show how time correlations in the fluctuating field correspond to different statistical sampling of the random measurements. Finally, we demonstrate for random telegraph noise \cite{RTN1,RTN2} the imprint of the time scale of the correlated noise on the final survival probability after applying the entire measurement sequence.

\section*{Model}
\subsection*{Stochastic Schrödinger Equation}
\begin{figure}[t]
\centering
 \includegraphics[width=0.95\linewidth]{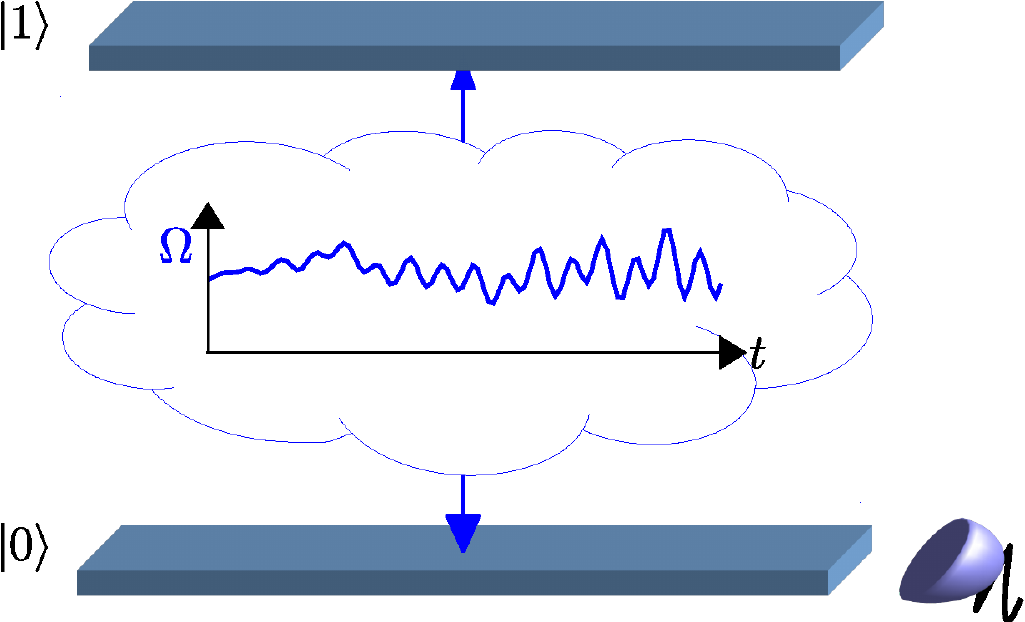}
 \caption{A random Hamiltonian driving couples the two levels $|0\rangle$ and $|1\rangle$ of a two-level system. The system is initially prepared in $|0\rangle$. By measuring the remaining population in $|0\rangle$, we can extract information about the fluctuating field driving the system dynamics.}
 \label{fig:overview}
\end{figure}

We study a quantum system that is coupled to a bath that effectively acts on the system via a time fluctuating classical field $\Omega(t)$ as
\be
H(t) = H_{0} + \Omega(t) H_{noise}= H_0 + [\overline{\Omega}+ \omega(t)] H_{noise} \; ,
\ee
where $H_{0}$ is the Hamiltonian of the unperturbed system, while $H_{noise}$ describes the coupling of the environment to the system. We assume that $\Omega(t)$ takes real values with mean $\overline{\Omega}$, and $\omega(t)$ is the fluctuating part with vanishing mean value. Figure~\ref{fig:overview} shows an exemplary two-level system initially prepared in the ground state $|0\rangle$. The random Hamiltonian driving term causes a population transfer to the upper level $|1\rangle$. This can be probed by measuring the remaining population in $|0\rangle$.
The system dynamics for a given realization of the random field $\Omega(t)$ is described by the standard Schrödinger equation. If we average over the statistics of the field $\Omega(t)$, we find the following master equation: 
\bea
\dot{\rho}(t)&=&-i[H_0+\overline{\Omega}H_{noise},\rho(t)]  \\
&+& \int_0^t \langle \omega(t)\omega(\tau) \rangle [H_{noise},[H_{noise},\rho(\tau)]] d\tau \; ,
\eea
where $\langle \omega(t)\omega(\tau) \rangle$ is the second-order time correlation of the random field, and $[\cdot,\cdot]$ represents the commutator. For white noise it is a Dirac delta distribution, i.e. $\langle \omega(t)\omega(\tau) \rangle\propto\delta(t-\tau)$, and we find the Lindblad-Kossakowski master equation~\cite{Kossakowski}. Otherwise, the memory kernel can lead to non-Markovian dynamics depending on the structure and time scale of the Hamiltonian as for example demonstrated for random telegraph noise (RTN) and $1/f$-noise~\cite{Benedetti1,Benedetti2}. Note that also a Markovian random field $\Omega(t)$ (as in the case of RTN) can lead to non-Markovian dynamics of the quantum system.

\begin{figure*}[t]
 \includegraphics[width=1.\linewidth]{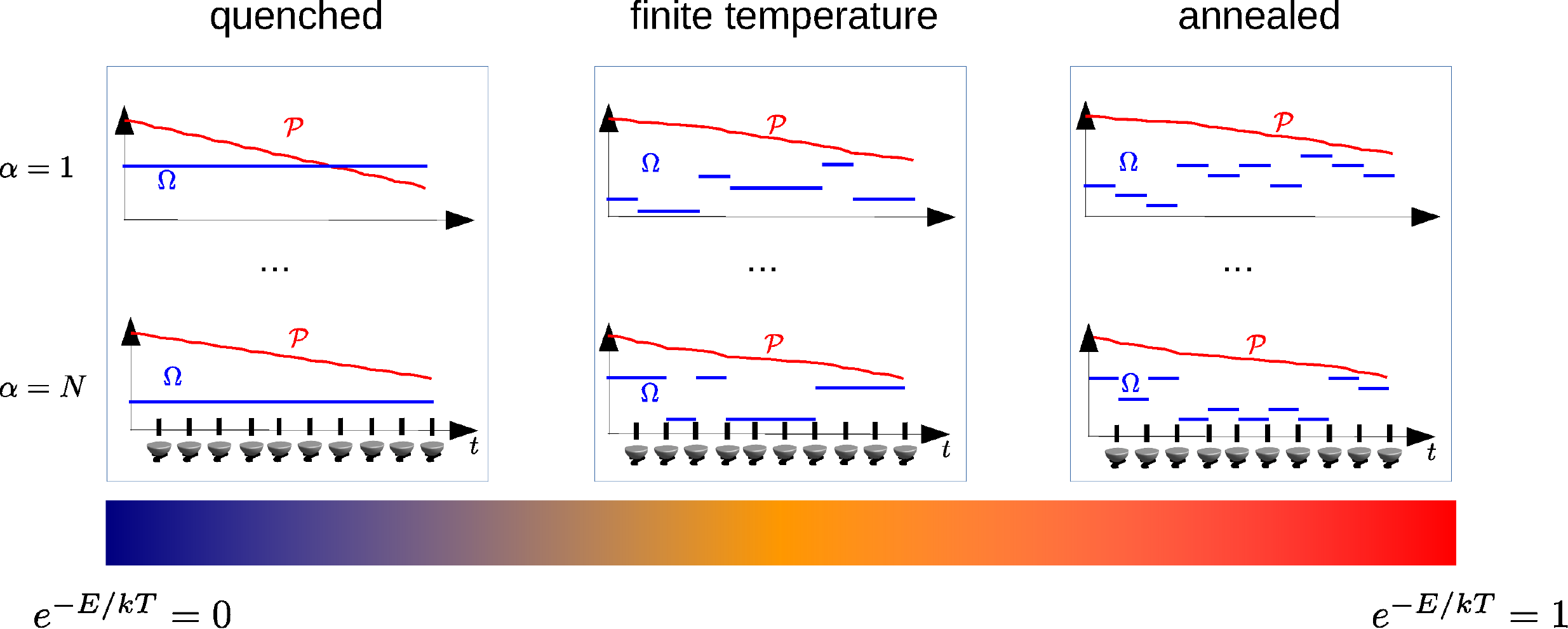}
\caption{Schematic view of the field fluctuations and their influence on the survival probability during the measurement sequence. The driving field $\Omega$ fluctuates in time and with increasing temperature the time correlations vanish going from quenched disorder to annealed disorder. The survival probabiltity $P$ decreases in time at a rate depending on the fluctuating value of the field. For annealed disorder the effect of the field fluctuations over a couple of time intervals is averaged out and for each realization $P$ converges to the same value. If we decrease the temperature the time correlation of the fluctuation grows and this convergence slows down. In the limit of $T=0$ the fluctuations degenerate to a random offset value that determines the behavior of $P$ that is now different for each realization.}
 \label{fig:sketch}
\end{figure*}

We now consider a system under sequential measurement where each measurement occurs after a fixed time interval $\mu$. We call $q(\Omega)$ the single measurement quantum survival probability that will depend on the value of $\Omega$ during this time interval and thus be a random variable. We can now generalize the survival probability to the stochastic process as follows
\be
P_{\alpha}(m)=\prod_{j=1}^{m}q(\Omega_{j,\alpha}) \; ,
\ee
where $\alpha=1,\dots N$ labels the realization of a trajectory, $j$ represents the time order of the $m$ measurements, and $\Omega_{j,\alpha}(t)$ is the fluctuating field in this corresponding time interval. To characterize it, two natural quantities arise:  the time average and the ensemble average of the survival probability.
The time-average is defined here as
\be\label{eq:time_average1}
\hat{P}_{\alpha}(m)\triangleq\lim_{M\rightarrow \infty}\frac{1}{M}\sum_{j=1}^{M} P_{\alpha}(j)^{\frac{m}{j}}.
\ee
The idea is that, using the measured value of the survival probability after the $j$-th measurement, one can estimate the expectation value at $m$ by $\hat{P}_{\alpha}(m)\approx P_{\alpha}(j)^{\frac{m}{j}}$. We can then average this value for $j=1\dots M$ and take the limit of a large number of measurements $M$. This limit potentially depends on the realization $\alpha$ of the fluctuating field as will be discussed below.
The ensemble-average is instead defined as
\be
\langle\mathcal{P}(m)\rangle\triangleq\lim_{N\rightarrow \infty}\frac{1}{N}\sum_{\alpha=1}^{N} P_{\alpha}(m),
\ee
where the average of $P_{\alpha}(m)$ is over a large number of realizations $N$. In the limit of infinite realizations this does not depend on the single realization but on their probability distribution.
In the following section, we examine the behaviours of the time and ensemble averages of the survival probability $P_{\alpha}(m)$, and in particular we study how correlations in the field fluctuations influence these averages.

\section*{Results}
For each realization $\alpha$ of the stochastic process we characterize the fluctuating field in between two measurements by a constant value  $\Omega_{j,\alpha}(t)\rightarrow \Omega_{j,\alpha}$ distributed according to a random distribution $p(\Omega)$. This is a valid formulation also for more complicated fluctuations, e.g. when the unitary dynamics is governed only by the fluctuating field, i.e. $H_0=0$. In the latter case, the single quantum survival probabilities to survive in the initial state $|\psi_0\rangle$ thus become
\be
q(\tilde\Omega)=|\langle\psi_0|\mathrm{e}^{-i \int_0^{\mu} \Omega(t) dt H_{noise} }|\psi_0\rangle|^2,
\ee
thus $q$ depends just on the constant $\tilde\Omega=\frac{1}{\mu}\int_0^{\mu}{\Omega}(t) dt$ with $\mu$ being the length of the time interval between two measurements. Note that for simplicity we chose the initial state $\rho_0=|\psi_0\rangle\langle\psi_0|$ to be pure. However, the main following results depend just on the statistics of $q(\Omega)$ and not on the actual dependence of $q$ on $\Omega$.
Thus, for non-vanishing $H_0$ we treat $\Omega$ just as a parameter that describes the statistics of $q(\Omega)$ via the distribution $p(\Omega)$.

Figure~\ref{fig:sketch} shows in the right upper panel how the fluctuating field $\Omega$ causes the survival probability $P$ to decrease at a fluctuating rate, i.e. a stronger average driving strength within one time interval causes a smaller $q$ and a faster decrease of $P$. Whitin each time interval between two measurements the decrease of $P$ is quadratic in the time interval and the field strength. While the field fluctuations are random, after a few measurements the influence of these fluctuations on $P$ are averaged out and the decay of $P$ behaves similarly for each realization.
When the field fluctuations are correlated, however, the decay of the survival probability depends much stronger on the realization because the probability distribution for $\Omega_{j+1,\alpha}$ depends on the value of $\Omega_{j,\alpha}$ (and potentially also on the previous history). This means that the convergence of the time average is much slower since a random deviation will influence not only a single time interval but a range of them, corresponding to the relaxation time associated to the time correlations.
Now, we consider a simple correlation model inspired by random telegraph noise (RTN)~\cite{RTN1,RTN2}: we choose $\Omega_{j+1,\alpha}$ according to the distribution $p(\Omega)$ only with a certain probability $\mathfrak{p}$, and $\Omega_{j+1,\alpha}=\Omega_{j,\alpha}$ otherwise. This update probability $\mathfrak{p}$ can be associated to a temperature $T$ by $\mathfrak{p}=\mathrm{e}^{-E/kT}$. The physical interpretation of this RTN environment is that the field value changes when for example a charge is trapped, and by thermal fluctuations a trapping energy barrier $E$ has to be overcome for the charge to be released such that the field value is restored to its previous value. In figure~\ref{fig:sketch} temperature grows from left to right yielding different types of disorder. For $T=0$, one has $\mathfrak{p}=0$, i.e. the value of the field $\Omega$ is chosen only once randomly and then always remains the same. The relaxation time is infinite and the time average does always converge to the same value. 
This scenario simulates the interaction of the system with an environment that exhibits \emph{quenched} disorder. Depending on the value of $\Omega$, the decay can be faster or slower. On the other side, for infinite temperature we have $\mathfrak{p}= 1$, representing an \emph{annealed} disorder environment. Between these two extreme regimes, i.e. for finite temperature, we have 
$\mathfrak{p}\in (0,1)$, hence a mixture of both behaviours. Here, \emph{quenched} disorder means a scenario with a static noise that depends on the initial random configuration of the environment, whereas \emph{annealed} disorder means that the environment changes its configuration randomly in time~\cite{Brout1,Emery1,Edwards1}.

\subsection*{Time and Ensemble Averages vs. Noise Correlations}
For the time average we introduce the expected frequencies $m\, n_{\Omega}$ with which the event $\Omega$ occurs in one realization of a stochastic sequence of $m$ measurements. The time average is then given by
\bea
\hat{P}_{\alpha}(m)=\lim_{M\rightarrow \infty}\frac{1}{M}\sum_{j=1}^{M} \prod_{\{\Omega\}} (q(\Omega)^{j n_{\Omega}})^{\frac{m}{j}}=\prod_{\{\Omega\}} q(\Omega)^{m\,n_{\Omega}}\,,
\eea
where the product is over all possible values of $\Omega$ and $n_{\Omega}$. For independent (thus uncorrelated) and identically distributed (i.i.d.) random variables $\Omega_{i,\alpha}$ the expected frequencies correspond to the underlying probability distribution $n_{\Omega}=p(\Omega)$. For correlated $\Omega_{i,\alpha}$ the convergence of the time average might not be unique or not even exist. This is linked to the Markov property and recurrence of the stochastic process~\cite{Lamperti}, as explained in more detail below. Note that a Markovian stochastic process $P_{\alpha}(m)$ does not imply Markovian quantum dynamics of the system since a Markovian fluctuating field can generate non-Markovian quantum dynamics through its time-correlations \cite{Benedetti1,Benedetti2}.

The ensemble average is instead the expectation value of the survival probability, i.e.
\bea
\label{eq:ensemble-average-annealed}
\langle \mathcal{P}(m)\rangle &=& \int d \mathcal{P} Prob(\mathcal{P})\mathcal{P}\\
&=& \int d \Omega_1\dots \int d \Omega_m  \prod_{i=1}^m p_i(\Omega_i|\Omega_1,\dots \Omega_{i-1})q(\Omega_i),\nonumber
\eea
where $Prob(\mathcal{P})$ is the probability distribution of the survival probability $\mathcal{P}$ (which is by itself a random variable depending on the field fluctuations) and $p_i(\Omega_i|\Omega_1,\dots \Omega_{i-1})$ is the conditional probability for the event $\Omega_i$ given the process history. In the case of i.i.d. random variables $\Omega_i$ it becomes
\bea
\langle \mathcal{P}(m)\rangle
&=&\int d \Omega_1\dots \int d \Omega_m  \prod_{i=1}^m p(\Omega_i)q(\Omega_i) \nonumber \\
&=&\left(\int p(\Omega) q(\Omega)\right)^m
\eea

\begin{figure}[t]
 \centering
 \includegraphics[width=1.\linewidth]{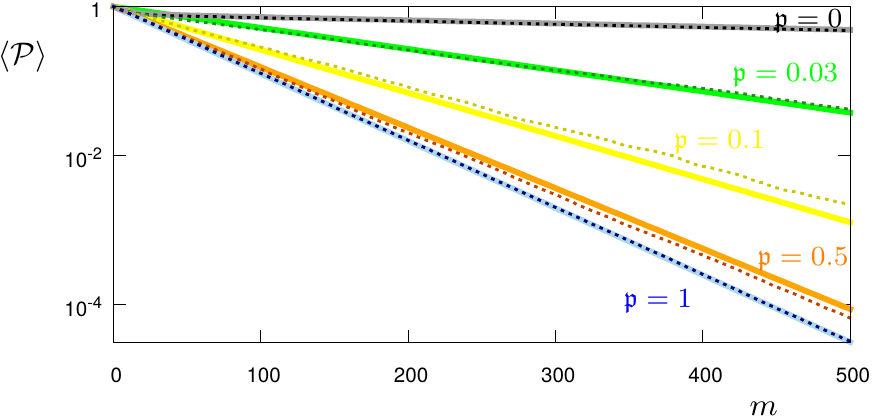}
 \caption{Ensemble Averages for $\mathfrak{p}=0,0.5,0.1,0.03,1$ (black, green, yellow, red, blue). The dashed lines correspond to the values calculated from 1000 realizations of the stochastic process, while the solid lines correspond to the respective theory curves.}
 \label{fig:ensemble-average}
\end{figure}

Now, we consider three different regimes to calculate the time and ensemble average by varying the value of $\mathfrak{p}$: 1) Annealed Disorder ($\mathfrak{p}=1$), 2) a finite temperature case with $\mathfrak{p}\in (0,1)$ and the number of measurements $m$ such that at least $5-10$ jumps occur, and 3) quenched disorder ($\mathfrak{p}=0$).
We introduce the shorthand notation $\langle A(\Omega)\rangle=\int p(\Omega)A(\Omega)d\Omega$ that we will use frequently for $A=q$ and $A=\ln q$.\\
In the case of annealed disorder (uncorrelated noise) the two averages follow straightforwardly from the definitions, namely
\be
\hat{P}_{\alpha}(m)_{an}=\exp\left\{m\langle\ln q(\Omega)\rangle\right\}
\ee
for the time average and
\be
\langle\mathcal{P}(m)\rangle_{an}=\exp\left\{m\ln\langle q(\Omega)\rangle \right\}
\ee
for the ensemble average -- see Eq.~(\ref{eq:ensemble-average-annealed}). In the case of quenched disorder, each single realization has constant $q(\Omega)$ and survival probability $q(\Omega)^m$. The ensemble average is the arithmetic average of these single possible outcomes, i.e.
\be
\langle\mathcal{P}(m)\rangle_{qu}=\exp\left\{\ln\langle q(\Omega)^m\rangle\right\}=\langle q(\Omega)^m\rangle \; .
\ee
Instead, the time average for quenched disorder does not take a single value but splits into several branches with
\be
\hat{P}_{\alpha}(m)_{qu}\in \{q(\Omega)^m\, |\,\Omega\in \mathrm{supp} (p(\Omega))\}
\ee
since the underlying process is not recurrent, in the sense that given the value of $\Omega$ in the first interval all the other values of the support of $p(\Omega)$ cannot be reached anymore within the given realization of the stochastic process.

Finally, for the finite temperature (fT) regime the problem is more difficult: The time average is the same as for annealed disorder, namely $\hat{P}_{\alpha}(m)_{fT}=\hat{P}_{\alpha}(m)_{an}$. The reason is that, despite of the time correlations, after each field update event the field is chosen according to $p(\Omega)$ independently from the history of the process. Also $\mathfrak{p}$ is independent of the current value of the field and for long times the frequencies of each $q(\Omega)$ converge to their expected values $n_{\Omega}=p(\Omega)$.
Instead, in order to calculate the ensemble average we have to take into account the correlations and examine the occurence of sequences of $\Omega(t)$ constant over several time intervals and updates of the probability according to $\mathfrak{p}$. If the length of such a sequence is labelled by $k$, this length $k$ is distributed by the Poisson distribution
$r_{\lambda}(k)=\frac{\lambda^k}{k!}\mathrm{e}^{-\lambda}$
with $\lambda=1/\mathfrak{p}$ the inverse of the probability for an update of $\Omega$. The survival probability for this sequence of constant field is:
\bea
\langle\mathcal{P}_{\mathfrak{p}}\rangle=\sum_{k=0}^{\infty}r_{\lambda}(k)\int p(\Omega) q(\Omega)^k d\Omega=\int p(\Omega)\mathrm{e}^{\frac{q(\Omega)-1}{\mathfrak{p}}} d\Omega \nonumber
\eea
The frequency of the updates is also Poisson distributed, with expectation value $\mathfrak{p}m$. The joint survival probability is then
\bea
\langle \mathcal{P}(m,\mathfrak{p})\rangle_{fT} &=& \mathrm{e}^{-\mathfrak{p}m}\sum_{k=0}^\infty \frac{(\mathfrak{p}m)^k}{k!} \langle\mathcal{P}_{\mathfrak{p}}\rangle^k \nonumber \\ 
&=&\exp\Big\{\mathfrak{p}m(\langle\mathcal{P}_{\mathfrak{p}}\rangle-1) \Big\}\,.
\eea
Fig.~\ref{fig:ensemble-average} shows the above calculated ensemble averages together with numerical values from the realization of $N=1000$ stochastic processes for different values of $\mathfrak{p}$. In all cases for $p(\Omega)$ we have used a bimodal distribution with $p_1=0.8$, $p_2=0.2$ and corresponding single measurement quantum survival probabilities $q_1=0.999$, $q_2=0.9$. If we decrease (increase) $q_1$ and $q_2$, the decay becomes faster (slower). The same happens for an increase (decrease) of $p_2$ that is the probability associated with $q_2<q_1$.

\subsubsection*{Variance of the Survival probability}
\begin{figure}[t]
\centering
  \includegraphics[width=1.\linewidth]{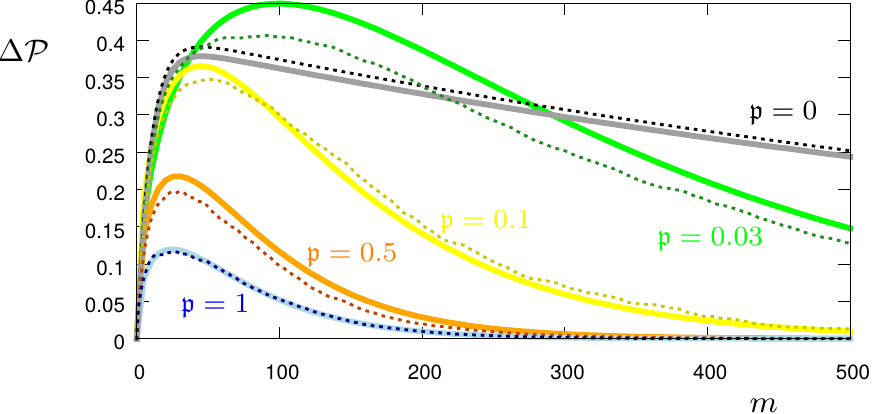}
 \caption{Standard Deviation for $\mathfrak{p}=0,0.5,0.1,0.03,1$ (black, green, yellow, red, blue). The dashed lines correpond to the value calculated from 1000 realizations of the stochastic process, while the solid lines correspond to the respective theory curve.}
 \label{fig:stdev}
\end{figure}
To calculate the variance of the distribution $Prob(\mathcal{P})$, we still need to calculate the second moment of the probability distribution of the survival probability. In the special case of infinite temperature or annealed disorder, it is given by
\bea
\langle \mathcal{P}^2(m)\rangle_{an} 
&=&\int d \Omega_1\dots \int d \Omega_m \prod_{i=1}^m p(\Omega_i)q(\Omega_i)^2
\nonumber\\
&=&\exp\left\{m\ln \langle q(\Omega)^2\rangle\right\}
\,.
\eea
The normalized variance thus reads
\bea
\frac{\Delta^2 \mathcal{P}(m)_{an}}{\langle \mathcal{P}(m)\rangle_{an}^2} &=&\frac{\langle \mathcal{P}(m)^2\rangle_{an}-\langle \mathcal{P}(m)\rangle_{an}^2}{\langle \mathcal{P}(m)\rangle_{an}^2}
\nonumber\\
&=&\exp\left\{m\left[ \ln \langle q(\Omega)^2\rangle - \ln\langle q(\Omega)\rangle^2 \right]\right\}-1. \; \; \; \; \;
\eea

For finite temperature, again we first consider a sequence of constant $\Omega$, where the square of the survival probability is as follows:
\bea
\langle\mathcal{P}^2_{\mathfrak{p}}\rangle &=& \sum_{k=0}^{\infty}r_{\lambda}(k)\int p(\Omega) (q(\Omega)^2)^k d\Omega \nonumber \\
&=& \int p(\Omega)\mathrm{e}^{\frac{q(\Omega)^2-1}{\mathfrak{p}}} d\Omega\,.
\eea
The frequency of field updates is again Poisson distributed, with expectation value $\mathfrak{p}m$. The joint squared survival probability is then
\bea
\langle\mathcal{P}(m,\mathfrak{p})^2\rangle_{fT} &=& \mathrm{e}^{-\mathfrak{p}m}\sum_{k=0}^\infty \frac{(\mathfrak{p}m)^k}{k!} \langle\mathcal{P}^2_{\mathfrak{p}}\rangle^k \nonumber\\
&=& \exp\Big\{\mathfrak{p}m(\langle\mathcal{P}^2_{\mathfrak{p}}\rangle-1) \Big\}\,,
\eea
and the normalized variance reads
\bea
\frac{\Delta^2 \mathcal{P}(m,\mathfrak{p})_{fT}}{\langle \mathcal{P}(m,\mathfrak{p})\rangle_{fT}^2}
=\exp\left\{\mathfrak{p}m\left[ \langle\mathcal{P}^2_{\mathfrak{p}}\rangle- 2\langle\mathcal{P}_{\mathfrak{p}}\rangle)+1\right]\right\}-1.  \; \; \; \;
\eea

Finally, for quenched disorder one has
\bea
\langle \mathcal{P}^2(m)\rangle_{qu}
=\int d\Omega p(\Omega)q(\Omega)^{2m}
=\exp\left\{\ln \langle q(\Omega)^{2m}\rangle\right\}, \; \; \; \; \; \; \; \;
\eea
with the normalized variance
\bea
\frac{\Delta^2 \mathcal{P}(m)_{qu}}{\langle \mathcal{P}(m)\rangle_{qu}^2}
=\exp\left\{\left[ \ln \langle q(\Omega)^{2m}\rangle - \ln\langle q(\Omega)^m\rangle^2 \right]\right\}-1. \; \; \; \; \; \; \; \; 
\eea

\begin{figure}[t]
\centering
  \includegraphics[width=1.\linewidth]{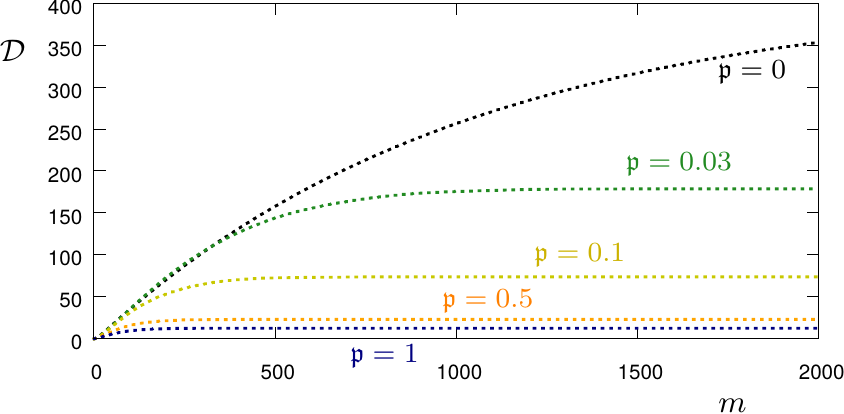}
 \caption{Accumulated standard deviation $\mathcal{D}(m) =\sum_{j=1}^m \Delta\mathcal{P}(j)$ for $\mathfrak{p}=0,0.5,0.1,0.03,1$ (black, green, yellow, red, blue). The dashed lines correpond to the values calculated from 1000 realizations of the stochastic process. For a relatively high number of measurements $m$ ($ > 300 $) there is a clear monotonicity of $\mathcal{D}$ as a function of the degree of the noise time-correlations related to the quantity $1-\mathfrak{p}$.}
 \label{fig:int-stdev}
\end{figure}

Figure~\ref{fig:stdev} shows the standard deviations (i.e. the square root of the above calculated variances, but without normalization) together with numerical values from the realization of 1000 stochastic processes for each chosen value of $\mathfrak{p}$. The underlying distribution $p(\Omega)$ is as in figure~\ref{fig:ensemble-average}. We find that the larger is the time-correlation (the smaller $\mathfrak{p}$), the larger is the standard deviation of the survival probability $\Delta\mathcal{P}$, i.e. the more the outcome depends on the single realization. To average out the non-monotonic bahaviour of $\Delta\mathcal{P}$, we consider the accumulated standard deviation $\mathcal{D}(m) =\sum_{j=1}^m \Delta\mathcal{P}(j)$, i.e. we sum up the standard deviation values for every measurement $j=1, \dots , m$. The result is shown in figure~\ref{fig:int-stdev}. Indeed, for relatively large values of $m$ ($ > 300 $) 
$\mathcal{D}(\mathfrak{p})$ does monotonically increase with the amount of noise temporal correlations related to the quantity $1-\mathfrak{p}$.

\section*{Discussion}
\begin{figure}[t]
\centering
\includegraphics[width=1.\linewidth]{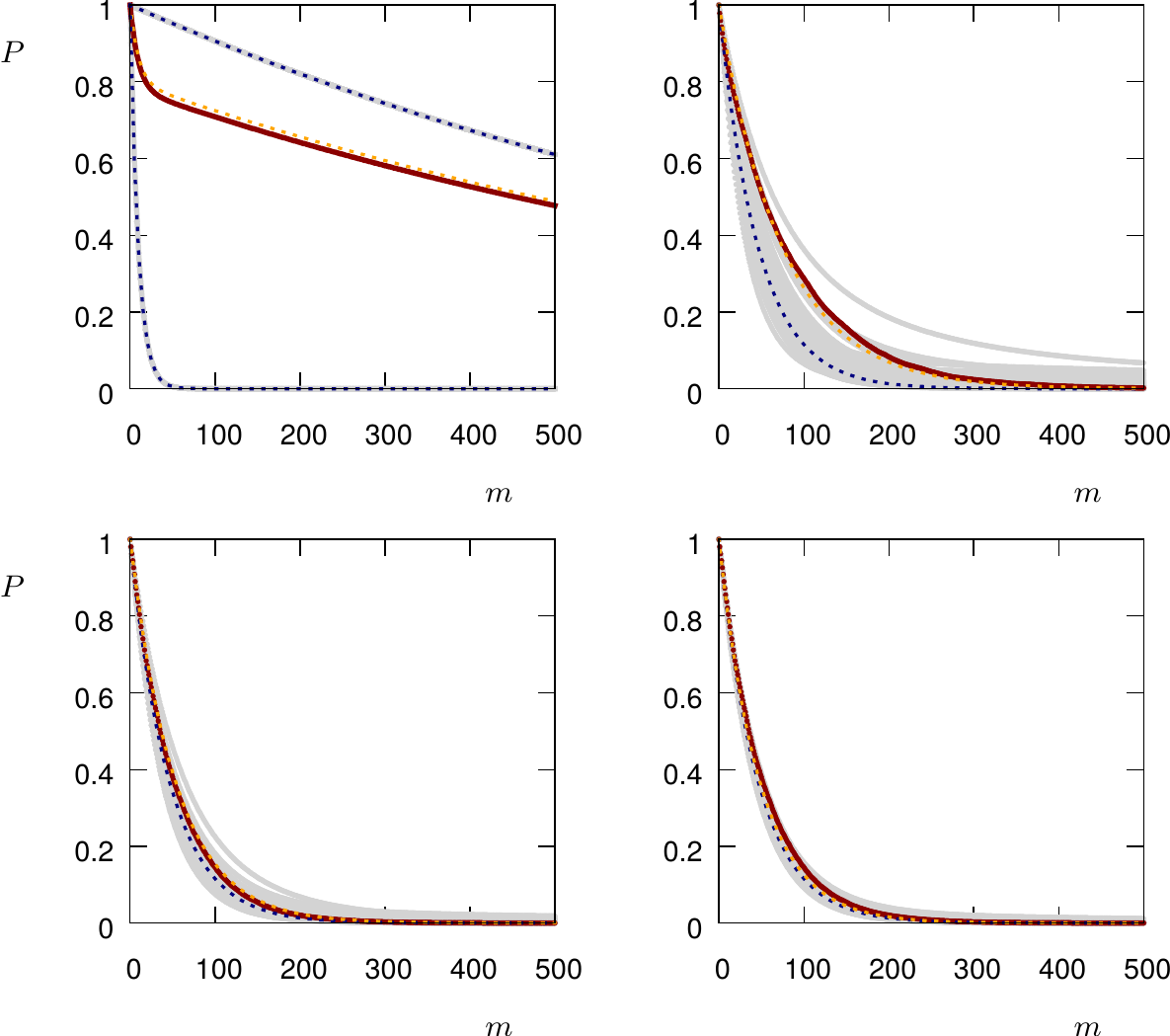}
\caption{Numerical Values: 50 realizations of the time average with $M=2000$ (grey solid lines), along with the ensemble average calculated from 1000 realizations of the stochastic process (red solid lines). These are compared to the theoretical curves for the time average (dark blue dashed) and ensemble average (orange dashed). Top left: quenched, top right: $\mathfrak{p}=0.1$, bottom left: $\mathfrak{p}=0.5$, bottom right: annealed.}
\label{fig:phasetransition}
\end{figure}

If we compare the time average with the ensemble one for different temperatures (i.e. $\mathfrak{p}$), we find that the convergence of the time average (quantified by the standard deviation) as well as the expected values for the ensemble average depend on $\mathfrak{p}$. If we consider $N$ realizations with $m$ measurements each, then for large numbers $m$ and $N$ (i.e. many measurements and many realizations) the frequency of each event $q(\Omega)$ is $mN p(\Omega)$ independently of $\mathfrak{p}$. To calculate the time average (for $\mathfrak{p}>0$) we will thus have $m p(\Omega)$ events and the value of the time average is then $\prod_{\{\Omega\}} q(\Omega)^{m p(\Omega)}$. For the ensemble average, instead, we have to average over many realizations, where each time the exponent of $q(\Omega)$ will deviate from $m p(\Omega)$ according to the (possibly time-correlated) statistics. Because of these increasing devations the ensemble average for annealed disorder is larger than the time average (as quantified in the appendix). If we include time correlations, the ensemble average will grow until it takes the maximum for the quenched disorder limit, i.e. the arithmetic average of the quantity $q(\Omega)^m$. As a consequence, we find
\begin{equation}\label{eq:averages-hierachy}
 \hat{P}_{\alpha}(m)_{an}\leq \langle\mathcal{P}(m)\rangle_{an}\leq \langle\mathcal{P}(m)\rangle_{fT}\leq \langle\mathcal{P}(m)\rangle_{qu} \; .
\end{equation}
For annealed disorder the time and ensemble averages practically coincide: we refer to this equality as an ergodic property of the system environment interaction~\cite{Gherardini2}. However, the more the $q(\Omega_{i,\alpha})$ are correlated, the more the ensemble average moves away from the time average and the ergodicity is broken. This can be seen in figure~\ref{fig:phasetransition} where time and ensemble averages are simulated for a bimodal distribution $p(\Omega)$ for quenched and annealed disorder, and for two values of finite temperature. In all cases, we have used a bimodal distribution with $p_1=0.8$, $p_2=0.2$ and corresponding single measurement quantum survival probabilities $q_1=0.999$, $q_2=0.9$. As shown in the appendix, the non-ergodic behavior depends essentially on the second and fourth moment of $p(\Omega)$. In other terms, for a similar average value, this effect will decrease if we choose $p_1\approx p_2$ or $q_1\approx q_2$. The same happens if we change the bimodal distribution into a multimodal or continuous distribution.
From an application point of view, this allows us to detect correlations in a fluctuating field by measuring and comparing to each other the time and ensemble averages of the survival probability. Furthermore, by changing the time interval $\mu$ between two measurements, we can explore the time scale on which these correlations occur.

\begin{figure}[t]
\centering
\includegraphics[width=1.\linewidth]{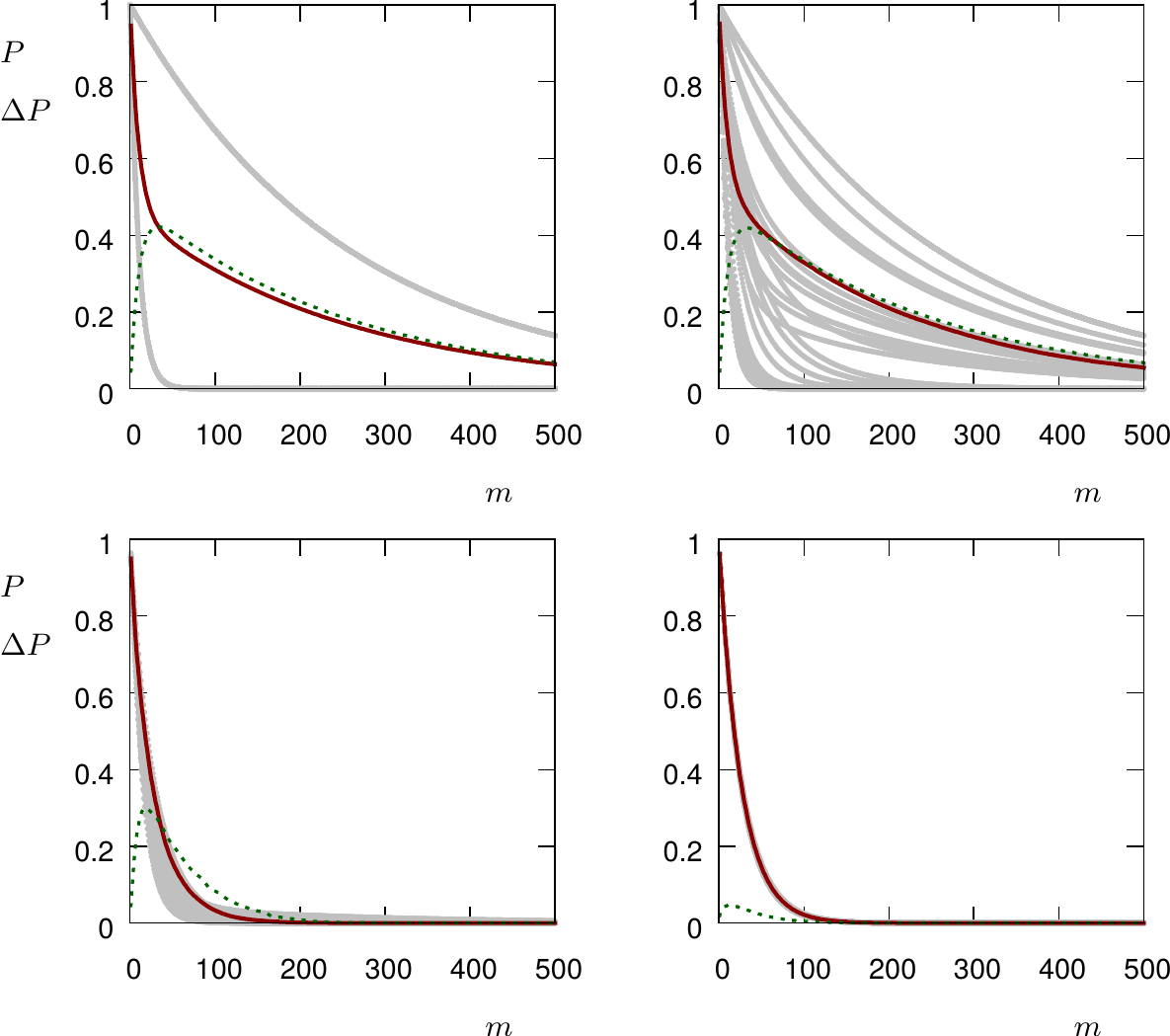}
\caption{Measurement sequence as in Figure  \ref{fig:phasetransition} but for the RTN qubit Hamiltonian described in the text. Numerical Values: 50 realizations of the time average (grey solid lines) with $M=2000$, along with the ensemble average calculated from 1000 realizations of the stochastic process (red solid line). The dark green dashed lines show the standard deviation of the single realizations. The time scale of the correlation decreases from left to right and from top to bottom, ranging from perfectly correlated (quenched) disorder to uncorrelated (annealed) noise.}
\label{fig:RTN}
\end{figure}

In order to test our method for a real quantum system, we consider the two-level Hamiltonian
\be
H=\Delta \ \sigma_z + \Omega(t)\sigma_x\,,
\ee
with the Pauli matrices $\sigma_x,\sigma_z$, a fluctuating driving $\Omega(t)$ of the system (e.g. an unstable classical light field) and a detuning term $\Delta$. We set $\Delta=2\pi\times 5\,$MHz and $\Omega\in 2\pi\times\{1,5\}$ MHz as a fluctuating RTN field with equal probability for both values. We initially prepare the system in the ground state $|0\rangle$ and perform projective measurements in this state spaced by intervals of constant length $\mu=100$ns. Such scheme may be implemented on many different experimental platforms and, very recently, has been realized in the stochastic quantum Zeno context with a Bose-Einstein condensate on an atom-chip~\cite{Gherardini1} under similar conditions.
Figure~\ref{fig:RTN} shows the time and ensemble averages along with the standard deviation, for an average time between the fluctuating field switches such as $10,10^3,10^5,10^7\,$ns. It can be clearly seen how a noise correlation time longer than the time interval $\mu$ generates a growing standard deviation of the survival probability, which can then be exploited as a witness of time-correlated noise.

\section*{Conclusions}
By studying the SQZD in time-correlated environments we have shown how an ergodicity property quantitatevely depends on the time scale of the noise correlations. By doing so, we propose a new (quantum Zeno-based) way to detect time correlations in random classical fields coupled to a quantum probing system. The time correlations in the noise field determine whether and how fast the survival probability converges to its statistical mean, hence tthe standard deviation of the survival probability over many experimental realizations will reveal information on the noise field. Then, we further improve this dependence by summing the standard deviation over the whole measurement series. Let us stress that this approach can be generalized by applying different measurement operators. Indeed, it has been demonstrated that quantum Zeno dynamics allows to confine the dynamics within decoherence-free subspaces \cite{Paz_Silva1,Maniscalco1}. By turning around that point of view, one can realize different initial states and measurement operators and thus also probe the effect of the environment on different subspaces of the system. Each of them might experience different time correlations of the noise as they predominantly couple to a different bandwidth of the noise spectrum. In conclusion we have introduced a novel method to examine time correlations and spectra of an environment acting on a quantum probing system as a random Hamiltonian term. This method does not rely on quantum state and process tomography but on a simple Zeno-based measurement scheme. It is also platform independent and thus can be used in very different implementations of the physical system and frequency/time scale ranges of the noisy environment field. Therefore, these results are expected to move further steps towards novel technologies for quantum sensing, where the fragile properties of quantum systems, as coherence, and especially here Zeno phenomena are exploited to probe an environmental fluctuating field and indirectly the presence of external artificial and biological molecules that are difficult to image otherwise.

\section*{Acknowledgements}
The authors gratefully acknowledge S. Ruffo, S. Gupta, A. Smerzi, and F.S. Cataliotti for useful discussions. This work was financially supported from the Ente Cassa di Risparmio di Firenze through the project Q-BIOSCAN.

\appendix
\section*{Appendix: Quantum Zeno regime}
If the time interval between two measurements $\mu$ is small compared to the system dynamics ($\Delta^2 H\mu^2\ll 1$) and we assume $H_0=0$, we can approximate $q(\tilde\Omega)$ by a Taylor expansion in $\tilde\Omega^2\mu^2$, where $\tilde{\Omega}=\frac{1}{\mu}\int_0^\mu \Omega(t) dt$ is again the mean value of the field within the time interval. For simplicity, from now on we obmit the tilde symbol and use $q(\Omega)$ and $\Omega$. Then, we consider also the Taylor expansion for the survival probability of the measurement sequence (note that this requires the stricter Zeno condition $m\Delta^2 H \mu^2\ll 1$). This allows us to analytically quantify the discrepancy between the different averages in the Zeno regime. In particular, we analyze the two extreme cases, i.e. the time average for annealed disorder and the ensemble average for quenched disorder.
%
%
Starting from Eq.~(\ref{eq:averages-hierachy}), we decrease the time interval $\mu$ such that we enter into the quantum Zeno dynamics regime. Then all time and ensemble averages collapse to one value (different from $1$, though). This approximation error is given by the quantity:
\begin{eqnarray*}
\frac{\langle\mathcal{P}(m)\rangle_{qu}-\hat{P}_{\alpha}(m)_{an}}{\hat{P}_{\alpha}(m)_{an}}
=\exp\left\{\ln\langle q(\Omega)^m\rangle-\langle \ln q(\Omega)^m\rangle\right\} - 1
 \nonumber \\ 
 \approx \Delta q \; ,
\end{eqnarray*}
with
\begin{eqnarray*}
\Delta q=\ln\langle q(\Omega)^m\rangle-\langle \ln q(\Omega)^m\rangle
=\frac{m^2}{2} (\Delta^2 H_{noise})^2(\omega_4-\omega_2^2) \mu^4 \; ,
\end{eqnarray*}
with $\omega_2$ and $\omega_4$ the second and third moment of the distribution $p(\Omega)$.
As a consequence, when we approach the Zeno limit all the averages collapse to the same value. Note that this is not trivial since the convergence is to the fourth order of the time interval while the leading order of the averages is the second one. The parameter $\Delta q$ can thus be considered to be the parameter that drives the transition from ergodic behavior to non-ergodic one, where in the strict Zeno regime all the effects of time correlations vanish regardless of the temperature. Instead, when we increase $\Delta q$ by moving out of the Zeno regime, temperature-dependent correlation effects cause a splitting of the values for the ensemble average of the survival probability.

\begin{figure}[b]
 \centering
 \includegraphics[width=1.\linewidth]{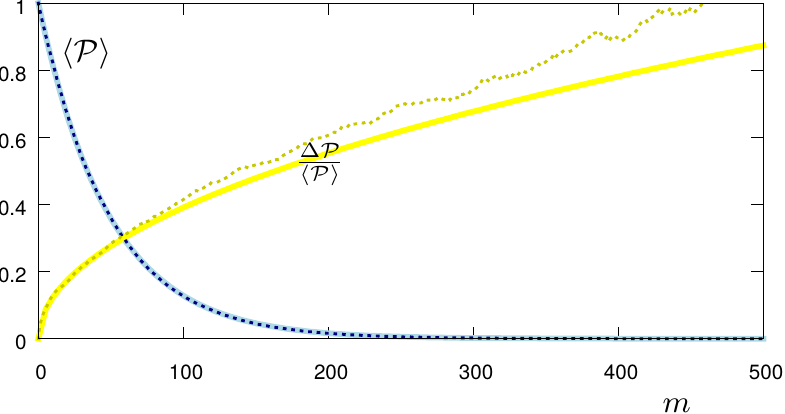}
 \caption{Ensemble average for 1000 realizations of a stochastic sequence with annealed disorder ($T=\infty$). $\langle\mathcal{P}(m)\rangle$ (simulation dark-blue, theory light-blue) and normalized standard deviation $\Delta \mathcal{P}(m) / \langle\mathcal{P}(m)\rangle$ (simulation dark-yellow, theory light-yellow) are shown as a function of $m$ .}
\label{fig:Annealed-Ensemble}
 \end{figure}

Concerning the variance of the distribution $Prob(\mathcal{P})$, in the special case of infinite temperature or annealed disorder, it is given by
\bea
\langle \mathcal{P}^2(m)\rangle_{an} &=& \int d \mathcal{P} Prob(\mathcal{P})\mathcal{P}^2 \nonumber\\ &=& \int d \Omega_1\dots \int d \Omega_m \prod_{i=1}^m p(\Omega_i)q(\Omega_i)^2
\nonumber\\
&=&\exp\left\{m\ln \langle q(\Omega)^2\rangle\right\}\,.
\eea
The normalized variance thus reads
\bea
\frac{\Delta^2 \mathcal{P}(m)_{an}}{\langle \mathcal{P}(m)\rangle_{an}^2} &=&\frac{\langle \mathcal{P}(m)^2\rangle_{an}-\langle \mathcal{P}(m)\rangle_{an}^2}{\langle \mathcal{P}(m)\rangle_{an}^2} \nonumber \\
&=&\exp\left\{m\left[ \ln \langle q(\Omega)^2\rangle - \ln\langle q(\Omega)\rangle^2 \right]\right\}-1 \nonumber\\
&\approx& m\left( \ln \langle q(\Omega)^2\rangle - \ln\langle q(\Omega)\rangle^2\right)\,,
\eea
which leads to the normalized standard deviation
\bea
\frac{\Delta \mathcal{P}(m)_{an}}{\langle \mathcal{P}(m)\rangle_{an}}
&\approx& \sqrt{m}\sqrt{\ln \langle q(\Omega)^2\rangle - \ln\langle q(\Omega)\rangle^2} \nonumber\\
&\approx& \sqrt{m}\Delta^2 H_{noise}\mu^2\sqrt{\omega_4-\omega_2^2}\,,
\eea
where the latter expression is a second-order expansion in the interval length, and $\omega_2$ and $\omega_4$ are the second and fourth statistical moments of $p(\Omega)$.
Fig.~\ref{fig:Annealed-Ensemble} shows the ensemble average along with the normalized standard deviation for annealed disorder as a function of $m$.

For the finite-temperature case, again we first consider a sequence of constant $\Omega$. The square of the survival probability is given by
\bea
\langle\mathcal{P}^2_{\mathfrak{p}}\rangle=\sum_{k=0}^{\infty}r_{\lambda}(k)\int p(\Omega) (q(\Omega)^2)^k d\Omega=\int p(\Omega)\mathrm{e}^{\frac{q(\Omega)^2-1}{\mathfrak{p}}} d\Omega. \nonumber
\eea
The frequency of the field updates is again Poisson distributed, with expectation value $\mathfrak{p}m$. The joint squared survival probability is then
\bea
\langle\mathcal{P}(m,\mathfrak{p})^2\rangle_{fT}=\mathrm{e}^{-\mathfrak{p}m}\sum_{k=0}^\infty \frac{(\mathfrak{p}m)^k}{k!} \langle\mathcal{P}^2_{\mathfrak{p}}\rangle^k
=\exp\Big\{\mathfrak{p}m(\langle\mathcal{P}^2_{\mathfrak{p}}\rangle-1) \Big\}. \nonumber
\eea
The normalized variance thus reads
\bea
\frac{\Delta^2 \mathcal{P}(m,\mathfrak{p})_{fT}}{\langle \mathcal{P}(m,\mathfrak{p})\rangle_{fT}^2}&=&\frac{\langle \mathcal{P}(m,\mathfrak{p})^2\rangle_{fT}-\langle \mathcal{P}(m,\mathfrak{p})\rangle_{fT}^2}{\langle \mathcal{P}(m,\mathfrak{p})\rangle^2}
\nonumber\\
&\approx&  m  \left[\frac{1}{\mathfrak{p}}+1 \right](\Delta^{2}H_{noise})^{2}\mu^4\omega_4\,,
\eea
which leads to the normalized standard deviation
\bea
\frac{\Delta \mathcal{P}(m,\mathfrak{p})_{fT}}{\langle \mathcal{P}(m,\mathfrak{p})\rangle_{fT}}
&\approx& \sqrt{m}\sqrt{1+\frac{1}{\mathfrak{p}}}\, \Delta^2 H_{noise}\mu^2\sqrt{\omega_4}\,.
\eea

For quenched disorder, one gets
\bea
\langle \mathcal{P}^2(m)\rangle_{qu} &=& \int d \mathcal{P} P(\mathcal{P})\mathcal{P}^2=\int d\Omega p(\Omega)q(\Omega)^{2m}  \nonumber \\
&=&\exp\left\{\ln \langle q(\Omega)^{2m}\rangle\right\}\,.
\eea
The normalized variance then reads
\bea
\frac{\Delta^2 \mathcal{P}(m)_{qu}}{\langle \mathcal{P}(m)\rangle_{qu}^2}&=&\frac{\langle \mathcal{P}(m)^2\rangle_{qu}-\langle \mathcal{P}(m)\rangle_{qu}^2}{\langle \mathcal{P}(m)\rangle_{qu}^2} \nonumber \\
&=&\exp\left\{\left[ \ln \langle q(\Omega)^{2m}\rangle - \ln\langle q(\Omega)^m\rangle^2 \right]\right\}-1   \nonumber \\
&\approx& \left( \ln \langle q(\Omega)^{2m}\rangle - \ln\langle q(\Omega)^m\rangle^2\right)\,,
\eea
and the normalized standard deviation is thus
\bea
\frac{\Delta \mathcal{P}(m)_{qu}}{\langle \mathcal{P}(m)\rangle_{qu}}
&\approx&  \sqrt{\ln \langle q(\Omega)^{2m}\rangle - \ln\langle q(\Omega)^m\rangle^2 }
 \nonumber \\
&\approx& m\Delta^2 H_{noise}\mu^2\sqrt{\omega_4-\omega_2^2}\,.
\eea
where the latter expression is a second-order expansion in the time interval length.

\bibliography{references}

\end{document}